\documentclass[intlimits,twoside,a4paper]{article}

\usepackage[cp1251]{inputenc}

\usepackage[eqsecnum]{cmpj3}

\issue{2026}{29}{3}{33701}
\doinumber{10.5488/CMP.29.33701}

\title[Ferromagnetic diatomic molecules]%
{Ferromagnetic diatomic molecules with mixed spin-1/2 and spin-1%
}
\author[E. Albayrak]{E. Albayrak\orcid{0000-0003-2695-0912}\thanks{Corresponding author: \email{albayrak@erciyes.edu.tr}}}
\address{
 Erciyes University, Physics Department, 38039 Kayseri, T\"{u}rk\.{i}ye
}
\date{Received 9 December 2025; revised 5 May 2026; accepted 5 May 2026; published 28 September 2026}
\begin{document}
	
\maketitle
\begin{abstract}
Diatomic molecules consisting of one spin-1/2 ($\sigma$) and one spin-1 ($S$) atoms are put at each site of the Bethe lattice (BL) of the coordination number $q=3, 4$, or 6. Alongside the ferromagnetic (FM) interactions of individual atoms within a molecule, the molecules are permitted to engage in ferromagnetic interactions with their nearest neighbors (NN) through diverse bilinear interaction parameters $J$. The spin-1 sites are also under the influence of crystal field $D$.  The phase diagrams on the ($D, T$) planes for a given $q$ are obtained by investigating the thermal variations of magnetizations for the case with zero external magnetic field $(H)$. It is discovered that the phase diagrams rely on $q$ both qualitatively and quantitatively. The crystal field is also included which operates only on the spin-1 sites. Three separate FM phases are observed with ($\sigma,S$) being $(1/2,1), (1/2,0)$ at the ground state level and a phase region with small magnetization (SM) values at a higher temperature range. In addition to the reentrant behavior in the transition zone from one phase to the other, the model produces first- and second-order phase transitions. Additionally, the impact of $H$ on the magnetization curves is examined, yielding quite intriguing findings.  
%
%
\printkeywords
%
\end{abstract}
\section{Introduction}

Spin models are often evaluated as a framework with a single spin at each lattice site, neglecting structural complexity and any internal interactions. Attributing a singular spin value to any composite material may be erroneous, as internal interactions, including the bilinear interaction parameter and crystal field, together with external factors such as temperature and external magnetic fields, can alter its behavior. As a result, it is possible to think of each lattice site as being occupied by two or more atoms, either of the same or different sorts. In addition, diatomic molecules are molecules composed of only two atoms, of the same or different chemical elements. If a diatomic molecule consists of two atoms of the same element, such as hydrogen (H$_2$) or oxygen (O$_2$), then it is said to be homonuclear. Otherwise, if a diatomic molecule consists of two different atoms, such as carbon monoxide (CO) or nitric oxide (NO), the molecule is said to be heteronuclear. 

Since the combined spin-1/2 and spin-1 model is the lowest conceivable mixing of spins and consequently probably the simplest, it has been the topic of numerous theoretical study. Mean-field approximations (MFA) were used in the hexagonal nanotube two-layer system \cite{Mendes}, the Ising-Heisenberg model on a honeycomb lattice with the effects of longitudinal and transverse crystal fields \cite{Albayrak}, and a three-dimensional Ising model with competing surface and bulk exchange interactions \cite{Benayad1,Benayad2}. Additionally, phase diagrams and magnetization curves of a magnetic superlattice \cite{Sarmento}, the impact of the crystal field on the honeycomb lattice \cite{Benyoussef1}, and a square lattice for different concentrations of magnetic atoms~\cite{Benyoussef2} were studied by using the effective-field theory (EFT). Additional studies under the EFT framework examine the diluted model on a honeycomb lattice subjected to a transverse field \cite{Kaneyoshi}, its critical behavior using the two-site cluster approach \cite{Bobakk}, the tricritical behavior on honeycomb and square lattices \cite{Lima}, the phase diagrams and magnetic properties with the uniaxial and biaxial single-ion anisotropy on a simple cubic lattice \cite{Belmamoun}, the magnetic behaviors of the ground state (GS) in a diluted system \cite{Xin}, as well as crystal-field interactions on honeycomb and square lattices \cite{Xin1} and \cite{Xin2}, respectively. A triangular lattice with two distinct spin-value distributions on the three sublattices \cite{Zukovic}, a plaquette of four-spin interaction on a two-dimensional square lattice \cite{Zaim}, the magnetic properties on a simple cubic lattice with crystal field \cite{Fouejio}, and a sublattice system with rectangular structures \cite{Obeidat} were all studied using Monte Carlo simulations. Phase diagrams in the presence of a crystal field \cite{Quadros}, the antiferromagnetic Ising model in two dimensions \cite{Boechat} and on a square lattice \cite{Boechat1} were obtained using renormalization-group techniques. Various other approaches were also carried out, such as the ones in terms of Oguchi approximation~\cite{Bobak,Bobak1}, in the EFT with correlations based on Glauber dynamics \cite{Keskin1}, the two-time Green-function technique \cite{JLi}, cluster variational theory within pair approximation \cite{Tucker}, a generalized star-triangle transformation~\cite{MJascur}, the double-time Green's function technique within the random phase decoupling approximation~\cite{Hu}, the pair approximation method~\cite{Boubekri} and by the transfer-matrix method~\cite{Lisnyi,Zad}. Similar works are also supplied by employing the exact recursion relations (ERR) to study the effects of a random crystal field on the phase diagrams~\cite{ErAl1} and to investigate the behaviors of the entropy and isothermal entropy change~\cite{ErAl2}.

It should be emphasized that the phrase ``heteronuclear diatomic molecules'' refers to the experimental realization of such diatomic molecules with two non-identical atoms. A molecule is generated by mixing atomic orbitals with various energies when atoms or spins are different. Each molecule orbit in the resulting polar bond receives an unbalanced contribution from atomic orbitals. Carbon monoxide in which both atoms are second-row elements, and hydrogen fluoride in which the two atoms are from different periods, are examples of heteronuclear diatomic molecules \cite{Gray}. Additionally, it has been established that the MnNi(EDTA)$\cdot$6H$_2$O complex is an example of a mixed-spin system \cite{Drillon}.

As can be observed in \cite{ALBAYRAKBL}, the ERRs were applied in the same manner in this study, i.e., single-spin at each site. Until recently it was the way to consider each site being inhabited by a single atom. Then, it was supposed that each site is occupied by a molecule of two atoms. Thermal fluctuations in magnetizations and magnetic phase diagrams of a diatomic molecule consisting of spin-1 atoms were investigated on the BL with the coordination number of $q=3$ by using the ERRs \cite{EA159}. Then it was extended to the investigation of diatomic molecules of either spin-1/2 and spin-1 \cite{EA164} or spin-1/2 and spin-3/2 \cite{EA176}, each of which is allowed to interact with its NNs. The diatomic molecules \cite{EA168} and triatomic molecules \cite{EA170} consisting of spin-1/2 atoms were also considered. The diatomic molecules consisting of spin-1/2 and spin-1 are again considered in this work for only the FM interactions between the atoms with various coordination numbers $q=3, 4$, and 6, and the thermal variations of magnetization of the systems are obtained to map the phase diagrams on the $(D, T)$ planes. The effect of the external magnetic field is also considered. 

This work is organized as follows: The next part covers the BL's molecular approach. The temperature fluctuations in the magnetizations of the atoms and the potential phase diagrams are shown in the third section, where some results are also discussed. 

\section{Formulation in terms of the ERRs}

\begin{figure}[!t]
	\centering{\includegraphics[width=0.5\textwidth]{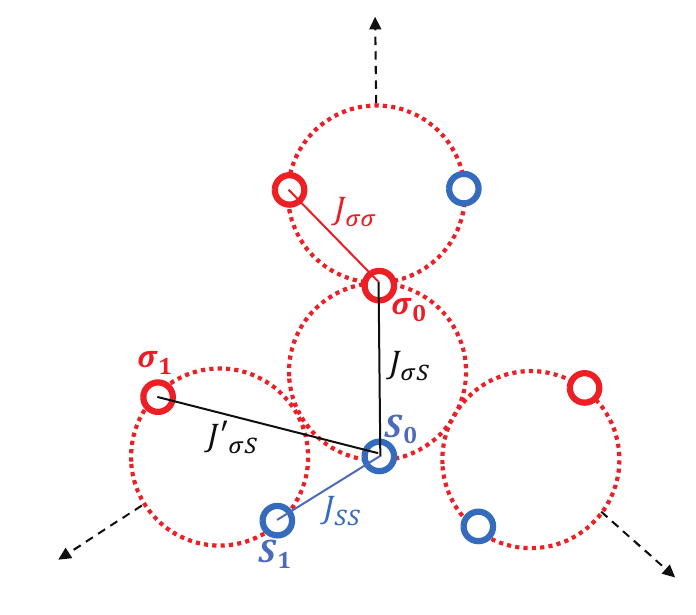}}
	\caption{(Colour online) The schematic representation of the BL in terms of the molecules and their atoms. Atoms are represented by solid circles, and molecules are represented by dotted circles. ($\sigma_0, S_0$) denotes the central molecule atoms, while ($\sigma_1, S_1$) are their NNs on the second shell and so on to infinity in the thermodynamic limit. The $J$ interactions between NN atoms are shown. The BL is only shown for the coordination number of three.}
	\label{fig1}
\end{figure}

The studied Hamiltonian of diatomic molecules having one spin-1/2, $\sigma$, and one spin-1 atom, $S$, can be written in terms of various bilinear interaction parameters $J$ by including the interactions between the atoms of each molecule, i.e., between spin-1/2 and spin-1, $J_{\sigma s}$, and between the atoms of NN molecules: $J_{ss}$ between spin-1 atoms, $J_{\sigma \sigma}$ between spin-1/2 atoms, and $J'_{\sigma s}$ between spin-1/2 and spin-1 atoms and the crystal field $D$ and external magnetic field $H$ acting only on the spin-1 atoms as
\begin{eqnarray} \label{eq:Hamiltonian}
	\mathcal{H}=&-& \sum_{\langle i,i\rangle} J_{\sigma s} \sigma_{i} S_{i}- \sum_{\langle i,j\rangle} J_{ss} S_{i} S_{j}
	- \sum_{\langle i,j\rangle}  J_{\sigma \sigma} \sigma_{i} \sigma_{j} \nonumber \\ &-& \sum_{\langle i,j\rangle}  J'_{\sigma s} ( \sigma_{i} S_{j}+S_{i} \sigma_{j})-D \sum_{i} S_{i}^2-H \sum_{i} (\sigma_{i}+S_{i}),
\end{eqnarray}
where the eigenvalues of spin operators $\sigma$ and $S$ are $\pm 1/2$ and $\pm 1$, 0, respectively, and $i, j$ count the sites of the BL and tend to infinity in the thermodynamic limit as shown in figure~\ref{fig1}. Furthermore, $\langle i,i \rangle$ denotes the interaction of atoms within the $i$th molecule, while $\langle i,j\rangle$ signifies the interaction between the NN molecules.  

The BL is built so that $q$ NN molecules of $\sigma_1$ and $S_1$, the initial shell molecules, interact with the center molecule, $i=0$, with $\sigma_0$ and $S_0$. Each molecule in the first shell is then connected to ($q-1$) molecules in the second shell, $\sigma_2$ and $S_2$. This routine is repeated until the thermodynamic limit is reached, i.e., the number of shells approaches infinity \cite{EA159,EA164}.

To acquire the ERRs of the model, one should start with the partition function given as: 
\begin{eqnarray}
	Z=\sum \re^{-\beta \hat{\mathcal{H}}}=\sum \limits_{Spc} P(Spc) ,
\end{eqnarray}
where the inverse temperature is $\beta=1/kT$, and $P(Spc)$ can be thought of as an unnormalized probability distribution over the spin configurations ($Spc$) \cite{Baxter}. It can be written for the central molecule as:  
\begin{equation}	
	P(\sigma_0, S_0)=\re^{\beta (J_{\sigma s} \sigma_0 S_0+D S_0^2)}  \prod_{j=1}^{q} Q_{n}\big(\{\sigma_0, S_0\}|\{\sigma_1, S_1\}^j\big),
\end{equation} 
where
\begin{eqnarray}	
	Q_{n}(\{\sigma_0, S_0\}|\{\sigma_1, S_1\}^j)&=&\re^{\beta (J_{\sigma s} \sigma_1 S_1+D S_1^2)}   \times \re^{\beta (J_{ss} S_0 S_1+J_{\sigma \sigma} \sigma_0 \sigma_1)}\nonumber \\  
	&\times&\re^{\beta J'_{\sigma s} (\sigma_0 S_1+ \sigma_1 S_0)} \times \re^{ {\scriptsize \text{Rest of the interactions}}}.
\end{eqnarray} 
It is obtained by severing the BL at the first shell, which comprises $q$ molecules, each of which has $q-1$ NN from the second shell. Thus, it is given as:  

\begin{eqnarray}	
	Q_{n}(\{\sigma_0, S_0\}|\{\sigma_1, S_1\})&=&\re^{\beta (J_{\sigma s} \sigma_1 S_1+D S_1^2)}    \times \re^{\beta (J_{ss} S_0 S_1+J_{\sigma \sigma} \sigma_0 \sigma_1)} \nonumber \\  
	&\times&\re^{\beta J'_{\sigma s} (\sigma_0 S_1+ \sigma_1 S_0)} \prod_{k=1}^{q-1} Q_{n-1}(\{\sigma_1, S_1\}|\{\sigma_2, S_2\}^k).
\end{eqnarray} 
This process is repeatedly employed until the thermodynamic limit is attained, which is accomplished by iterations until the computations are stabilized. 

In order to get the ERRs, one needs to calculate equations (3)--(6) over the possible spin configurations in the same form as for the central molecule
\begin{equation}
	g_{n}(\{\sigma_0, S_0\})= \sum\limits_{\{\sigma_1, S_1\}} Q_{n}\big(\{\sigma_0, S_0\}|\{\sigma_1, S_1\}\big)
\end{equation}
and for the first shell molecules 
\begin{equation}
	g_{n-1}(\{\sigma_1, S_1\})= \sum\limits_{\{\sigma_2, S_2\}} Q_{n-1}\big(\{\sigma_1, S_1\}|\{\sigma_2, S_2\}\big)
\end{equation}
in which the first is computed for the values of $\sigma_0$ and $S_0$ and then summing over $\sigma_1$ and $S_1$, while the second is calculated for the given values of $\sigma_1$ and $S_1$ and then summing over for $\sigma_2, S_2$. Since the diatomic molecules can have one spin-1/2 with two possible eigenvalues, $\pm1/2$, and one spin-1 with three possible eigenvalues, $\pm1,0$, then one gets six $g_{n}(\{\sigma, S\})$ functions, which yield five ERRs for each diatomic molecule. They are found from 
\begin{eqnarray}	
	&&X_1=\frac{g_n(\frac{1}{2},1)}{g_n(-\frac{1}{2},-1)}, \quad X_2=\frac{g_n(\frac{1}{2}, 0)}{g_n(-\frac{1}{2},-1)}, \quad X_3=\frac{g_n(\frac{1}{2},-1)}{g_n(-\frac{1}{2},-1)}, \nonumber \\  && \hspace{1.5cm}
	X_4=\frac{g_n(-\frac{1}{2},1)}{g_n(-\frac{1}{2},-1)}, \quad X_5=\frac{g_n(-\frac{1}{2}, 0)}{g_n(-\frac{1}{2},-1)},
\end{eqnarray}	
which are given explicitly as:
\begin{eqnarray}	
	X_1&=& \big[\re^{\beta(D+0.25J_{\sigma \sigma}+0.5J_{\sigma s}+J'_{\sigma s}+J_{ss}+1.5H)}X_1^p+
	\re^{\beta(0.25J_{\sigma \sigma}+0.5J'_{\sigma s}+0.5H)}X_2^p\nonumber \\ 	
&	+&\re^{\beta(D+0.25J_{\sigma \sigma}-0.5J_{\sigma s}-J_{ss}-0.5H)}X_3^p+
	\re^{\beta(D-0.25J_{\sigma \sigma}-0.5J_{\sigma s}+J_{ss}+0.5H)}X_4^p\nonumber \\ 
	&+&\re^{\beta(-0.25J_{\sigma \sigma}-0.5J'_{\sigma s}-0.5H)}X_5^p+
	\re^{\beta(D-0.25J_{\sigma \sigma}+0.5J_{\sigma s}-J'_{\sigma s}-J_{ss} -1.5H)}\big]/X, \nonumber
\end{eqnarray}
\begin{eqnarray}	
	X_2&=&[\re^{\beta(D+0.25J_{\sigma\sigma}+0.5J_{\sigma s}+0.5J’_{\sigma s}+1.5H)}X_1^p +
	\re^{\beta(0.25J_{\sigma\sigma}+0.5H)}X_2^p \nonumber \\ 
	&+&
	\re^{\beta(D+0.25J_{\sigma\sigma}-0.5J_{\sigma s}-0.5J’_{\sigma s}-0.5H)}X_3^p +
	\re^{\beta(D-0.25J_{\sigma\sigma}-0.5J_{\sigma s}+0.5J’_{\sigma s}+0.5H)}X_4^p \nonumber \\ 
	&+&
	\re^{\beta(-0.25J_{\sigma\sigma}-0.5H)}X_5^p +
	\re^{\beta(D-0.25J_{\sigma\sigma}+0.5J_{\sigma s}-0.5J’_{\sigma s}-1.5H)}]/X, \nonumber
\end{eqnarray}
\begin{eqnarray}	
	X_3&=&[\re^{\beta(D+0.25J_{\sigma\sigma}+0.5J_{\sigma s}-J_{ss}+1.5H)}X_1^p +
	\re^{\beta(0.25J_{\sigma\sigma}-0.5J’_{\sigma s}+0.5H)}X_2^p \nonumber \\ 
	&+&
	\re^{\beta(D+0.25J_{\sigma\sigma}-0.5J_{\sigma s}-J’_{\sigma s}+J_{ss}-0.5H)}X_3^p +
	\re^{\beta(D-0.25J_{\sigma\sigma}-0.5J_{\sigma s}+J’_{\sigma s}-J_{ss}+0.5H)}X_4^p 
	\nonumber \\ 
	&+&
	\re^{\beta(0.5J’_{\sigma s}-0.25J_{\sigma\sigma}-0.5H)}X_5^p +
	\re^{\beta(D-0.25J_{\sigma\sigma}+0.5J_{\sigma s}+J_{ss}-1.5H)}]/X, \nonumber
\end{eqnarray}
\begin{eqnarray}	
	X_4&=&[ \re^{\beta(D-0.25J_{\sigma\sigma}+0.5J_{\sigma s}+J_{ss}+1.5H)}X_1^p +
	\re^{\beta(0.5J’_{\sigma s}-0.25J_{\sigma\sigma}+0.5H)}X_2^p \nonumber \\ 
	&+&
	\re^{\beta(D-0.25J_{\sigma\sigma}-0.5J_{\sigma s}+J’_{\sigma s}-J_{ss}-0.5H)}X_3^p +
	\re^{\beta(D+0.25J_{\sigma\sigma}-0.5J_{\sigma s}-J’_{\sigma s}+J_{ss}+0.5H)}X_4^p \nonumber \\ 
	&+&
	\re^{\beta(+0.25J_{\sigma\sigma}-0.5J’_{\sigma s}-0.5H)}X_5^p +
	\re^{\beta(D+0.25J_{\sigma\sigma}+0.5J_{\sigma s}-J_{ss}-1.5H)}]/X, \nonumber
\end{eqnarray}
\begin{eqnarray}	
	X_5&=&[\re^{ (\beta(D-0.25J_{\sigma\sigma}+0.5J_{\sigma s}-0.5J’_{\sigma s}+1.5H)}X_1^p +
	\re^{\beta(-0.25J_{\sigma\sigma}+0.5H)}X_2^p \nonumber \\ 
	&+&
	\re^{\beta(D-0.25J_{\sigma\sigma}-0.5J_{\sigma s}+0.5J’_{\sigma s}-0.5H)}X_3^p +
	\re^{\beta(D+0.25J_{\sigma\sigma}-0.5J_{\sigma s}-0.5J’_{\sigma s}+0.5H)}X_4^p \nonumber \\ 
	&+&
	\re^{\beta(0.25J_{\sigma\sigma}-0.5H)}X_5^p +
	\re^{\beta(D+0.25J_{\sigma\sigma}+0.5J_{\sigma s}+0.5J’_{\sigma s}-1.5H)}]/X, \nonumber
\end{eqnarray}
with
\begin{eqnarray}	
	X&=&\re^{\beta(D-0.25J_{\sigma\sigma}+0.5J_{\sigma s}-J’_{\sigma s}-J_{ss}+1.5H)}X_1^p+
	\re^{\beta(-0.25J_{\sigma\sigma}-0.5J’_{\sigma s}+0.5H)}X_2^p\nonumber \\ 
	&+&
	\re^{\beta(D-0.25J_{\sigma\sigma}-0.5J_{\sigma s}+J_{ss}-0.5H)}X_3^p+
	\re^{\beta(D+0.25J_{\sigma\sigma}-0.5J_{\sigma s}-J_{ss}+0.5H)}X_4^p\nonumber \\ 
	&+&
	\re^{\beta(0.25J_{\sigma\sigma}+0.5J’_{\sigma s}-0.5H)}X_5^p+
	\re^{\beta(D+0.25J_{\sigma\sigma}+0.5J_{\sigma s}+J’_{\sigma s}+J_{ss}-1.5H)}. \nonumber
\end{eqnarray}
where $p=q-1$.

The partition function, which is a key component of practically any statistical technique, can be obtained from the definition given as
\begin{eqnarray}
	&&Z=\sum_{ \{\sigma_0, S_0\}}
	\re^{\beta [J_{\sigma S} \sigma_0 S_0+D S_0^2+H(\sigma_0 +S_0)]}  [g_n( \sigma_0, S_0)]^q.
\end{eqnarray} 
and then it is calculated in terms of the ERRs by using the eigenvalues of the $\sigma$ and $S$ spins as 
\begin{eqnarray}
	Z =\Big[g_n\big(-\frac{1}{2},-1 \big)\Big]^q\times Z' 
\end{eqnarray}
which will be used in obtaining the magnetizations. Here,
\begin{eqnarray}
	Z'&=& \big[\re^{\beta(D+0.5J_{\sigma s}+1.5H)}X_1^q+\re^{0.5\beta H}X_2^q \nonumber \\&+& \re^{\beta(D-0.5J_{\sigma s}-0.5H)}X_3^q+\re^{\beta(D-0.5J_{\sigma s}+0.5H)}X_4^q+\re^{-0.5\beta H}X_5^q \nonumber \\
	&+&
	\re^{\beta(D+0.5J_{\sigma s}-1.5H)}\big].
\end{eqnarray}

We are now able to calculate the magnetizations of each atom for spin-1/2 as
\begin{eqnarray}
	M_{\sigma}
	&=&\frac{1}{Z}\sum_{ \{\sigma_0, S_0\}} \sigma_0
	\re^{\beta [J_{\sigma S} \sigma_0 S_0+D S_0^2+H(\sigma_0 +S_0)]}  [g_n( \sigma_0, S_0)]^q \nonumber \\
	&=& \frac{1}{2Z'}[\re^{\beta(D+0.5J_{\sigma s}+1.5H)}X_1^q+\re^{0.5\beta H}X_2^q+
	\re^{\beta(D-0.5J_{\sigma s}-0.5H)}X_3^q\nonumber \\
	&-&\re^{\beta(D-0.5J_{\sigma s}+0.5H)}X_4^q-\re^{-0.5\beta H}X_5^q -\re^{\beta(D+0.5J_{\sigma s}-1.5H)}]
\end{eqnarray} 
and for spin-1 as 
\begin{eqnarray}
	M_{S}
	&=&\frac{1}{Z}\sum_{ \{\sigma_0, S_0\}} S_0
	\re^{\beta \big[J_{\sigma S} \sigma_0 S_0+D S_0^2+H(\sigma_0 +S_0)\big]}  [g_n( \sigma_0, S_0)]^q \nonumber \\
	&=& \frac{1}{Z'} \big[\re^{\beta(D+0.5J_{\sigma s}+1.5H)}X_1^q-\re^{\beta(D-0.5J_{\sigma s}-0.5H)}X_3^q\nonumber \\
	&+&\re^{\beta(D-0.5J_{\sigma s}+0.5H)}X_4^q-\re^{\beta(D+0.5J_{\sigma s}-1.5H)}\big]
\end{eqnarray} 
and thus the average magnetization of a diatomic molecule is just given as  
\begin{eqnarray}
	M_{\sigma S}=\frac{1}{2} [M_{\sigma}+M_{S}].
\end{eqnarray} 

After obtaining all of the formulations of the model in terms of ERRs, we can now compute its variations for given values of $J$ and $D$ under temperature change to derive various phase diagrams of the model.

\section{The results and conclusions}
 Before presenting our illustrations, a few words are in order about the nature of parameters. Since we are dealing with the spins only along one axis, call it simply the $z$-axis. The thermal agitations ($T$) drive the model into the paramagnetic (P) ordering, since the spins are randomly oriented along the $z$-axis. $J>0.0$ aligns the spins along the $z$-axis in the FM ordering. The spins have the tendency to align with the $H$ which is along the $z$-axis. Since $D$ is only effective at spin-1 sites, its positive values drive the spins to the GS values of $\pm1$, while its negative enough values pick the 0 GS value. Therefore, it is in a very critical role in the model since it affects the behaviors of phase transition properties.  
\begin{figure}[!t]
	\centering{\includegraphics[width=0.8\textwidth]{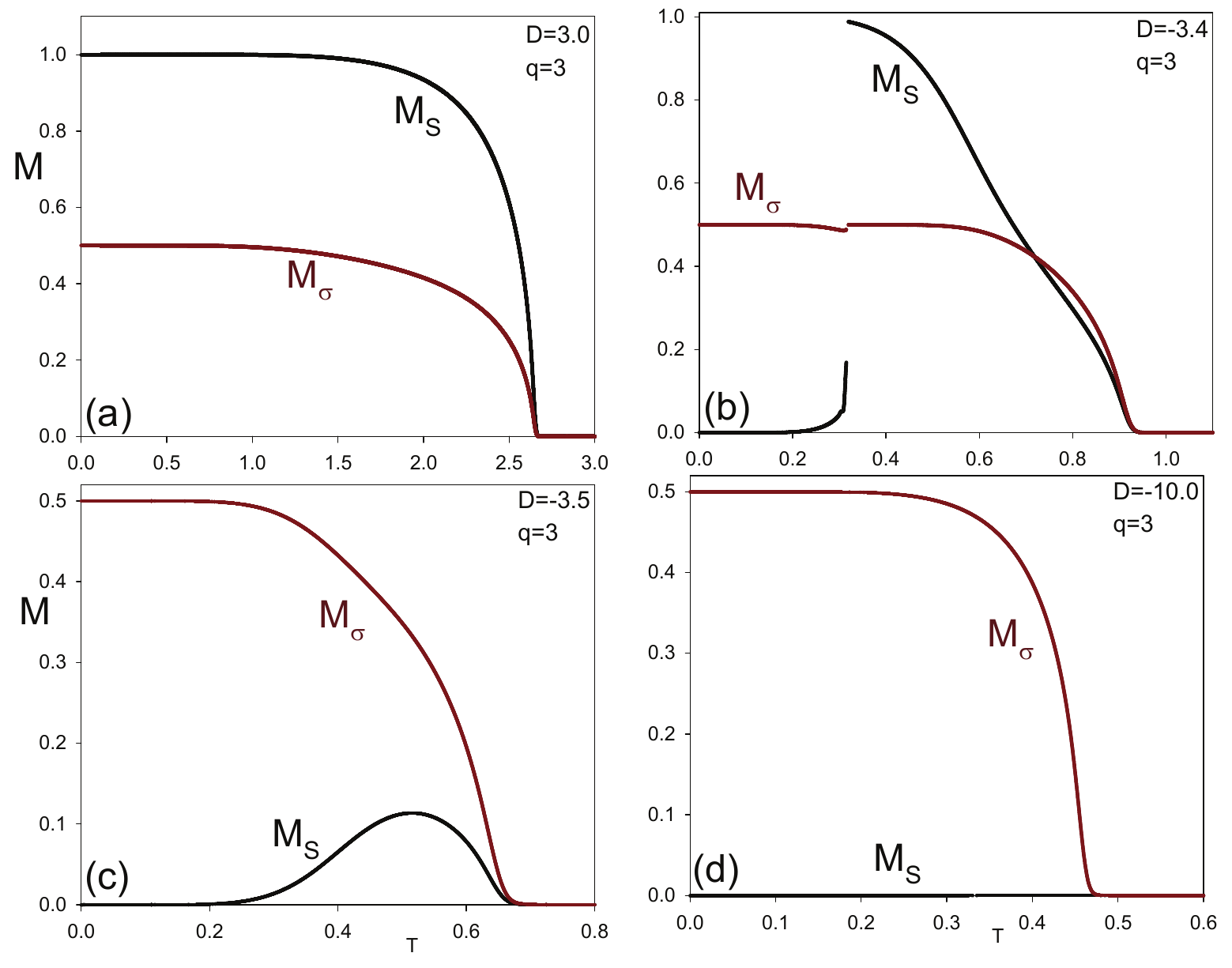}}
	\caption{(Colour online) Thermal variations of magnetizations $M_{\sigma}$ and $M_{S}$. They are obtained for all $J$ values set to 1.0 to represent FM interactions when $H=0.0$ for given $D$ as (a) 3.0, (b) $-3.4$, (c) $-3.5$, and (d) $-10.0$ for $q=3$.}
	\label{fig2}
\end{figure}
\begin{figure}[!t]
	\centering{\includegraphics[width=0.8\textwidth]{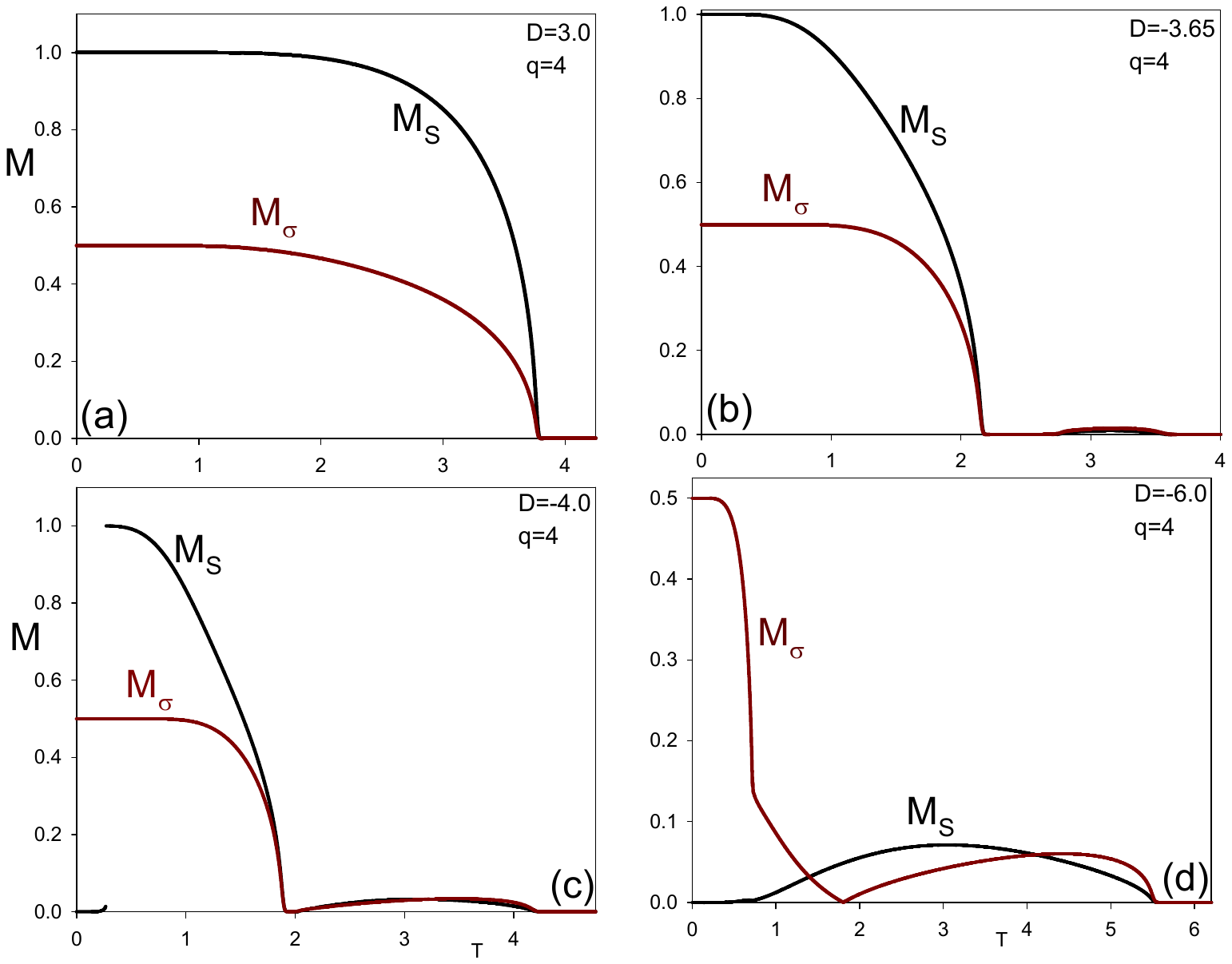}}
	\caption{(Colour online) Thermal variations of magnetizations $M_{\sigma}$ and $M_{S}$. They are obtained for all $J$ values set to 1.0 to represent FM interactions when $H=0.0$ for given $D$ as (a) 3.0, (b) $-3.65$, (c) $-4.0$, and (d) $-6.0$ for $q=4$.}
		\label{fig3}
\end{figure}
\begin{figure}[!t]
	\centering{\includegraphics[width=0.8\textwidth]{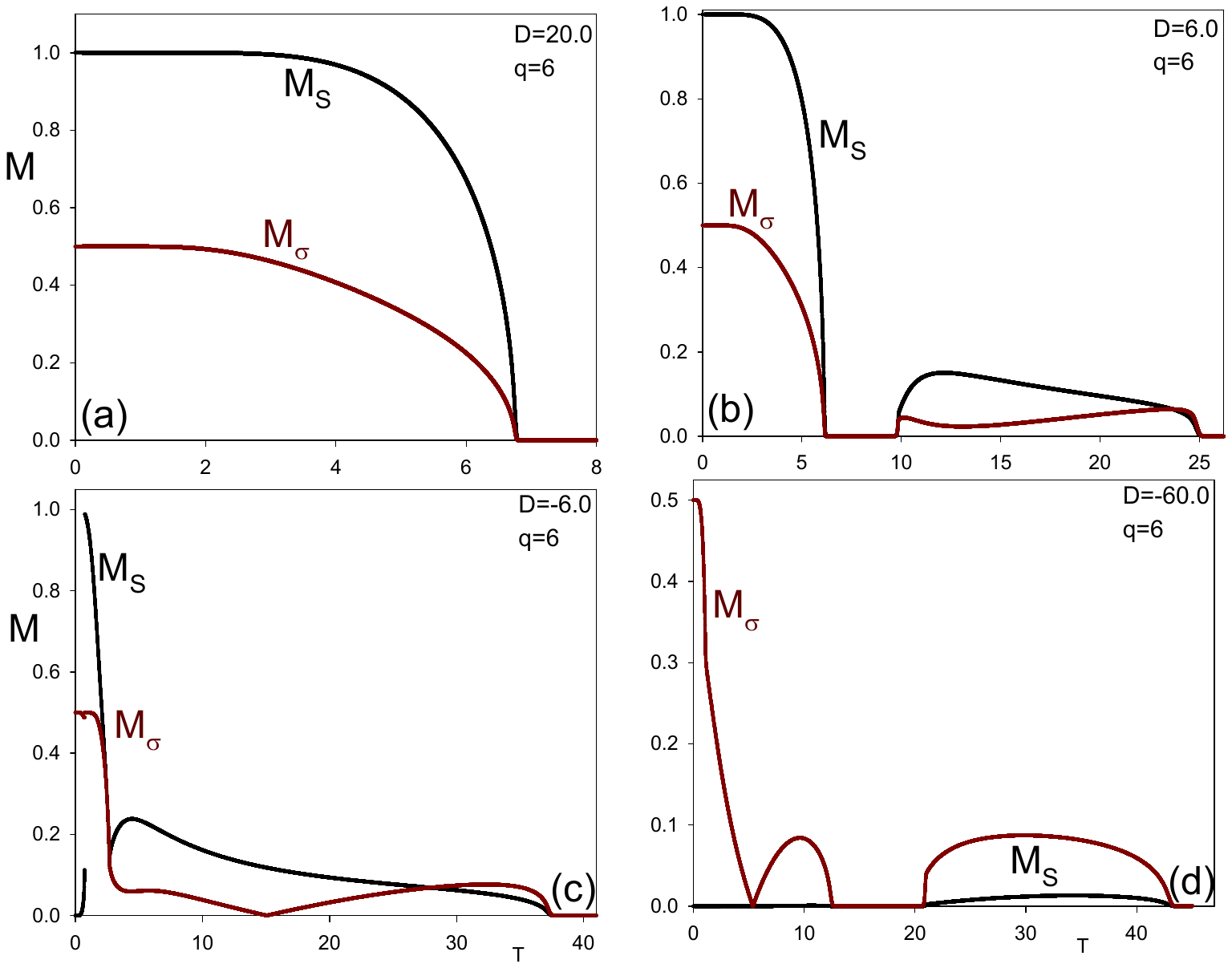}}
	\caption{(Colour online) Thermal variations of magnetizations $M_{\sigma}$ and $M_{S}$. They are obtained for all $J$ values set to 1.0 to represent FM interactions when $H=0.0$ for given $D$ as (a) 20.0, (b) 6.0, (c) $-6.0$, and (d) $-60.0$ for $q=6$.}
		\label{fig4}
\end{figure}
First we illustrate some characteristic magnetization curves obtained when $q=3, 4,$ and 6 with the given values of $D$ and all $J=1.0$ corresponding to the FM interaction when $H=0.0$. They are given in figures~\ref{fig2}(a)--(d), figures~\ref{fig3}(a)--(d), and figures 4(a)--(d) for $q=3, 4$, and 6, respectively. Figure~\ref{fig2}(a), figure~\ref{fig3}(a), and figure~\ref{fig4}(a) show that $M_{S}$'s start from 1.0 while $M_{\sigma}$'s start from 1/2 as expected for the $D>0$ when $T=0.0$; thus, in total, we see the behavior of a spin-3/2 model. As the temperature increases, magnetizations decrease to enter the P phase. The transition between the FM and P phases is continuous and therefore it is called the second-order phase transition, seen at a temperature $T_c$. The magnetizations are seen at higher temperatures as $q$ gets higher. When $D=-3.4$, see figure~\ref{fig2}(b), one observes $M_{S}=0.0$ and $M_{\sigma}=1/2$ for $q=3$ at the GS level. As $T$ increases, $M_{S}$ gains some value and then presents a discontinuous jump at the first-order phase transition temperature $T_t$ towards the value of 1.0, where $M_{\sigma}$ also presents a small jump.  As $T$ increases further, $M_{\sigma}$ and $M_{S}$ intersect with each other. Then they diminish at the common $T_c$. Figure~\ref{fig2}(c) is similar to figure~\ref{fig2}(b), but now the $T_t$ is not seen; instead, $M_{S}$ starts from zero, gains some value with increasing $T$, and then disappears at the $T_c$ continuously as $M_{\sigma}$.  In figure~\ref{fig2}(d), we see that $M_{S}$ is always zero as it is supposed to, since $D=-10.0$ is large enough to drive spin-1 to its zero GS value. In figure~\ref{fig3}(b) for $q=4$, the magnetizations start as in that of figure~\ref{fig3}(a) and terminate at the $T_c$, but then they start to gain some value as $T$ increases, albeit they are small, and with further increase, they disappear again. Therefore, the FM-P-FM-P type transitions appear. The small magnetization regions start with the $T_c$ and terminate with the $T_c$ again. Therefore, all together one sees three $T_c$s. In figure~\ref{fig3}(c), one also observes a visible jump at the $T_t$ in $M_S$ where $M_{\sigma}$ presents a very tiny jump at low $T$ in addition to the behaviors observed in figure~\ref{fig3}(b). It should also be noted that in figure~\ref{fig3}(c), the low temperature phase transition is of the first-order which accompanies the phases given as $(1/2,0)\rightarrow(1/2,1)$. In figure~\ref{fig3}(d), the magnetizations initiate at the GS values of $(1/2,0)$. As temperature $T$ rises, $M_{S}$ consistently increases, while $M_{\sigma}$ initially decreases, approaching zero (albeit small, it is not zero, i.e., it is not a phase transition point) with increasing $T$, before subsequently rising alongside $M_{S}$. Both magnetizations converge to zero at the critical temperature $T_c$.
For $q=6$, figure~\ref{fig4}(b) shows the same behavior obtained in figure~\ref{fig3}(b) for $q=4$. This time phase regions are clearer and seen at higher $T$, since $q$ is higher. Figure~\ref{fig4}(c) first presents a $T_t$ at low $T$, then the P phase is reached as in figure~\ref{fig3}(d) of $q=4$ again. Note also that $M_{\sigma}$ decreases again, approaching zero as in figure~\ref{fig3}(d); even though very small, it does not disappear; therefore, it is not a phase transition point. Figure~\ref{fig4}(d) shows that $M_{S}$ starts from zero, gains very small values around the first $T_c$, and then tends to zero along with $M_{\sigma}$ at the first $T_c$. Both magnetizations become zero, and a further increase in $T$ leads to the appearance of both magnetizations at the second $T_c$, and a further increase leads them to be diminished again at the third $T_c$. It is clear from these magnetization curves that the characteristic behaviors of magnetizations are similar for each $q$, but the temperature of the critical temperatures is seen at higher temperatures as mentioned before.  
\begin{figure}[!t]
	\centering{\includegraphics[width=0.7\textwidth]{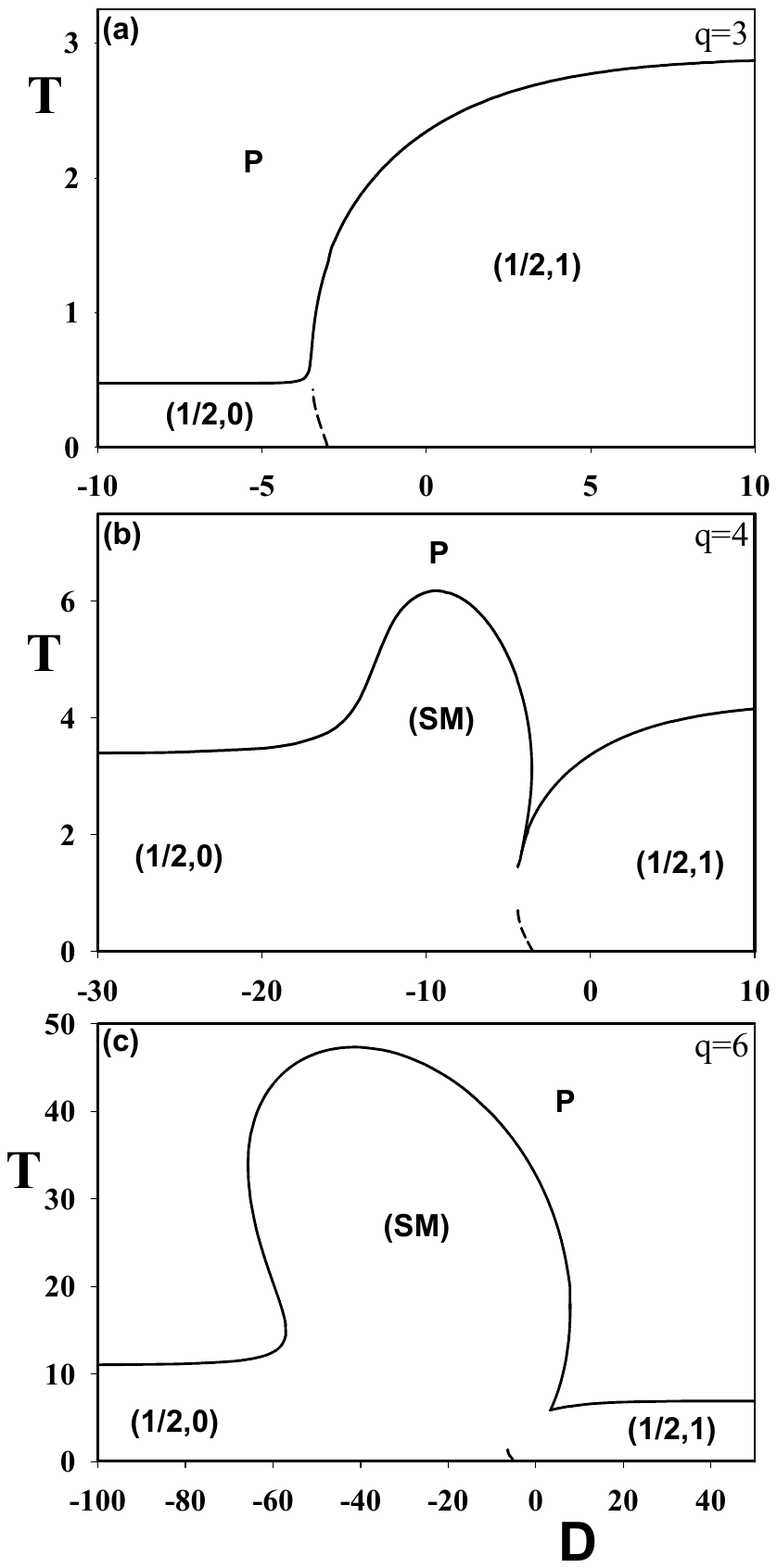}}
	\caption{The phase diagrams on the $(D, T)$ planes with all $J=1.0$ and $H=0.0$. The solid and dashed lines correspond to $T_c$ and $T_t$ lines, respectively. 
		They are obtained for a given $q$: (a) 3, (b) 4, and (c) 6.}
		\label{fig5}
\end{figure}

The detailed analysis of the magnetization curves when $H=0.0$ leads to the phase diagrams on the~$(D,T)$ planes with all $J=1.0$ when $q=3, 4,$ and 6 as shown in figures~\ref{fig5}(a)--(c). In figure~\ref{fig5}(a), for $q=3$, a phase diagram similar to the phase diagram of spin-3/2 \cite{Keskin} is obtained but with different phase regions. The phase is the FM $(1/2,0)$ phase at low negative $D$ values. As $D$ is increased towards positive values, $M_{S}$ starts gaining some values; therefore, the $T_c$ temperatures increase, and one also observes a $T_t$-line separating $(1/2,0)$ and $(1/2,1)$ FM phases. The latter phase adds up to give a phase region of 3/2 of spin-3/2, which is very logical.

When $q=4$, figure~\ref{fig5}(b), one sees again for high enough $D$, i.e., the right-hand side of the figure, the FM $(1/2,1)$ phase is the evidence giving a total of phase 3/2 again. The $T_t$-line is now shifted to the left, and the critical temperatures are higher in comparison to figure~\ref{fig5}(a). The $T_c$-line decreases as $D$ decreases; after making a dip, it instantly starts rising in the phase region of SM, where SM means the region of very small magnetization values. The FM phase region increases, and then after making a peak, starts decreasing to take the phase $(1/2,0)$ again. The last phase diagram obtained for $q=6$ and shown in figure~\ref{fig5}(c) is similar to figure~\ref{fig5}(b), with critical temperatures seen further up high and the $T_t$-line shifted further left. The story of the $T_c$-lines for $q=4$ and 6 is as follows: When $D$ is high enough towards its positive values, our system acts like spin-3/2; then as $D$ gets smaller, the inclination of $M_{S}$ preference for the zero spin state of spin-1 intensifies as $D$ becomes increasingly negative. This lowers the $T_c$ values. Therefore, as $D$ becomes more negative, the control of the system is left to spin-1/2 atoms. Thus, when $M_{S}$ suddenly becomes very small, the interaction of spin-1/2 atoms with all its NNs causes a rise in $T_c$. As $D$ gets further negative, $M_{S}$ now totally disappears, and the $T_c$-line becomes constant, leaving the phase with (1/2,0). It should be noted that the drastic change when one goes from $q=3$ to higher $q$ is actually needs detailed arguments.   
\begin{figure}[!t]
	\centering{\includegraphics[width=0.8\textwidth]{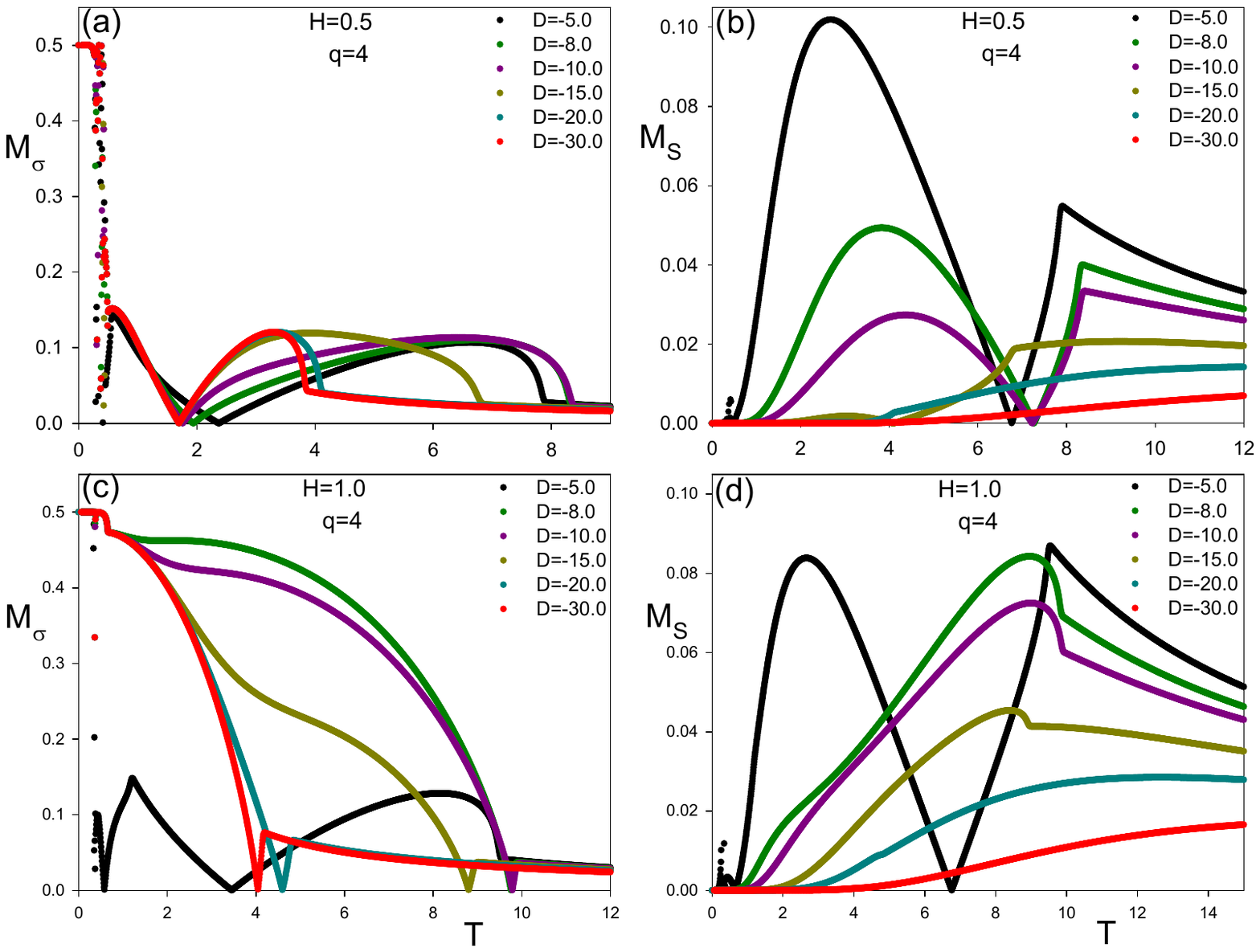}}
	\caption{(Colour online) Thermal variations of magnetizations $M_{\sigma}$ and $M_{S}$ when $H\neq0.0$. (a) and (c) show $M_{\sigma}$, and (b) and (d) show $M_{S}$. They are obtained for all $J$ being 1.0, (a)--(b) for $H=0.5$ and (c)--(d) for $H$=1.0 when $D=-5,-8,-10,-15,-20$, and $-30.0$ for all for and $q=4$ only.}
		\label{fig6}
\end{figure}   

As final illustrations, the magnetization curves are graphed for given values of $D<0.0$ when $H\neq0.0$ and $q=4$. Figure~\ref{fig6}(a)--(b) and figure~\ref{fig6}(c)--(d) are obtained for $H=0.5$ and $H=1.0$, respectively. These curves correspond to the SM regions. $M_{\sigma}$ starts from 1/2, while $M_{S}$ starts from zero at $T=0.0$. There are some peaked regions first, and then they tend very slowly and continuously to zero as expected when $H\neq0.0$. The dips are seen at lower $T$ for higher $H$. The jumps are clear at low $T$, indicating first-order transitions. The continuous peaks also shift to the left for higher $D$. Again, $D$, thus, the appearance of the zero state of spin-1 is responsible for all these dips and peaks. The appearance of peaks after the dip is again caused by the $J$ values, since each atom interacts with its three NNs and spin-1/2 gets the control of the model.

The present work has revealed new results for $q=4$ and 6. As a final remark, it should be noted that the SM phase is unique in that it manifests at extremely high temperatures with low magnetization values and a second-order phase transition into the PM phase. As far as our literature search is concerned, we were not able to find any study of diatomic molecules containing spin-1/2 and spin-1 for comparison reasons.

\ukrainianpart

\title{Феромагнітні двоатомні молекули зі змішаними спінами 1/2 та 1}
\author{
	Е. Албайрак
}
\address{
Університет Ерджіес, фізичний факультет, 38039 Кайсері, Туреччина
}

\makeukrtitle

\begin{abstract}
	\tolerance=3000%
		Двоатомні молекули, що складаються з одного атома зі спіном 1/2 ($\sigma$) та одного атома зі спіном 1 ($S$), розміщені в кожному вузлі ґратки Бете з координаційним числом $q=3, 4$ або 6. Поряд з феромагнітними взаємодіями окремих атомів у молекулі, останні можуть взаємодіяти з найближчими сусідами через різноманітні білінійні параметри взаємодії $J$. Центри зі спіном 1 також знаходяться під впливом кристалічного поля $D$. Фазові діаграми на площинах ($D, T$) для заданого $q$ отримані шляхом дослідження теплових змін намагніченості при нульовому зовнішньому магнітному полем $(H)$. Виявлено, що фазові діаграми залежать від $q$ як якісно, так і кількісно. Також включено кристалічне поле, яке діє лише на центрах зі спіном 1. Спостерігаються три окремі феромагнітні фази, де ($\sigma,S$) становлять $(1/2,1), (1/2,0)$ на рівні основного стану, а фазова область з малими значеннями намагніченості існує у діапазоні вищих температур. Окрім реентрантної поведінки в перехідній зоні від однієї фази до іншої, модель також демонструє фазові переходи першого та другого роду. Крім того, досліджується вплив $H$ на криві намагнічування, що показує досить цікаві результати.
	\keywords двоатомна молекула, змішаний спін, ґратка Бете, рекурсійні співвідношення, фазова діаграма
	
\end{abstract}

\end{document}